# ¿ The physical reality of the quantum wave function ?

Gerd Leuchs
Max Planck Institute for the Science of Light, Erlangen
Department of Physics, University of Erlangen-Nuremberg
Department of Physics, University of Ottawa

The quantum theory is amazingly successful and we all know how to use it. For a given experimental situation we can predict the outcome of the measurement. Often the prediction tells us the probability with which the detector will give a certain result. When dealing with a single quantum system allowing for several simultaneous measurements on sub-systems, then quantum physics can make precise predictions about correlations of the measured data if the system is in an entangled quantum state. David Mermin described the situation as follows: *"The only proper subjects of physics are correlations among different parts of the physical world. Correlations are fundamental, irreducible, and objective. They constitute the full content of physical reality"* [1]. The numerous discussions on the topic are well reflected in his American Journal of Physics article [2, see also 3]. One particular question is what happens in the process of measuring a quantum system, when quantum state reduction takes place. In his wonderful essay 'On quantum theory' Berge Englert stresses that there is no measurement problem. One by one he goes through all the different topics, some of us find cumbersome.  With regard to the measurement process he writes: "*... the collapse of the state" or "wave function collapse" are popular synonyms for state reduction. The connotation that the transition* (associated with state reduction) *is a dramatic dynamical process, as if the physical system were evolving, is clearly misleading"* [4]. And indeed it would be quite troublesome if a single photon or even a single massive particle going through a beam splitter would simultaneously exit through both output ports, only to "collapse" upon a measurement at one or the other output port [5]. This would be troublesome because energy or mass would have to move instantaneously over finite distances. But not being allowed to ask what happens inside the apparatus is also somewhat cumbersome, at least for experimenters. Therefore, attempts were made to avoid this cumbersome aspect by searching for alternative interpretations of quantum physics, but then a cumbersome aspect is bound to pop up somewhere else - there does not seem to be a free lunch.

| INTERPRETATION | remark | wave function | you should not ask |
|---|---|---|---|
| Copenhagen | knowledge collapses, not the wave function | not real | what happens inside the apparatus? |
| many worlds | different worlds evolve by measurement | real | where are all these worlds, exponentially growing in number? |
| Ghirardi-Rimini-Weber | measurement collapses the wave function | real | for the physical dynamics of wave function collapse (?) |
| QBism [6] | uses Bayesian probabilities | not real | how can nature be properly described by personalized probability amplitudes? |

<u>*Table 1:*</u>  **There is no free lunch**.
*This list of interpretations is not complete and just shows examples. The different interpretations listed here agree in the predictions of experimental results. But all interpretations have their cumbersome aspects, some find difficult to accept. These cumbersome aspects are merely moved around. This teaches us to accept a cumbersome aspect, one or the other way.*

Some scientists find the interpretation most appealing, which interprets the wave function and the quantum physical predictions as statements about the information we have about a system rather than as descriptions of the physical reality [4]. Thus there may be a deep underlying connection between quantum physics and information theory [see e.g. 5, 7, 8]. Maybe the long lasting discussion about interpretations tells us something. The different combinations of theory and interpretation seem to be equally valid although the different interpretations by scientists seem to be contradicting. So maybe the link between quantum physics and information theory is deeper than some of us think.

The experimenter will not be able to provide evidence for one interpretation being correct and the others having limited applicability as long as the different interpretations - or rather the theories behind them - do not lead to contradicting predictions for the outcome of an experiment. So it is no surprise that again and again experiments are performed to keep quantum theory on the test bench. In the following a selection of experiments providing such tests are reviewed and a new experiments are proposed.

## Violation of Bell's inequalities

Regarding a possible reality associated with the quantum wave function there had been ample speculations about the existence of hidden variables. Early on Heisenberg argued qualitatively, that hidden variables couldn't exist, because the quantum classical boundary is not well defined but depends on our ability to control the system (see [9] p. 506, 507). Then John Bell proposed his celebrated inequality to quantitatively test for local hidden variables [10], requiring systems in entangled quantum states, leading to a series of subsequent measurements ever improving with regard to potential loop holes, see e.g. [11-16]. Today it is clear that there is less and less room for local hidden variables in quantum theory. We note in passing, that John Bell's inequality is currently studied also in the classical regime [17], i.e. in the context of classically excited non-separable mode functions, such as cylindrical vector beams [18-21]. It may be too early to make final remarks but the violation may be associated with classical correlations being arbitrarily strong though not non-local. In any case applications in metrology were already proposed [22].

## Wheeler's delayed choice experiment

Coming back to the single photon impinging on a beam splitter, one has a choice to either measure at which output port the photon appears, or one can let the two output beams interfere at a second beam splitter closing the geometry to form an interferometer and then measure the interference of the photon. Clearly, in either case only one detector will fire. Without the recombining beam splitter both detectors will fire stochastically, the relative percentage reflecting the splitting ratio. However, with the beam recombiner and for the appropriate path difference the photon will always exit at one output port of the recombining beam-splitter, or for a properly modified phase change at the other output port and only one detector will fire the other one being silent [23]. In this latter case the photon obviously[1] seems to exit via both outputs of the first beam splitter in order for the wave function to interfere via the two paths. Needless to say, both these different experimental scenarios and the corresponding measurement results are properly described by quantum theory. Yet it is meaningless to ask whether the photon is really split or not, unless one goes and looks, i.e. measures in the two interferometer arms – but then one is back to the first scenario. John Archibald Wheeler had the idea to refine the apparatus and decide whether or not to insert the beam recombiner only after the photon has passed the first beam splitter [24]. This proposal is feasible if the photon travel distances are so large that the associated travel times are longer than the time constant of the switch with which one alternates between the which-path (i.e. 'welcher-Weg') and the interference measurement. The result is that one sees no difference as compared to the static cases. Switching back and forth before the photon arrives does not have any influence on the measurement results [25-28]. Further studies of the topic use weak measurements in order to gain more information [7, 23, 29] and the discussion about it [30].

---

[1] To quote the experimenter

## The "01 – 10" state

Little more than ten years ago there was a fierce debate about whether the quantum state of the single photon is entangled or not after passing through the beam splitter. There is no simpler wave function than that of a single photon underlining the relevance of this discussion. The opponents argued that by a mere transformation to other modes the state, which looked like displaying entanglement between a single photon state and the vacuum ground state, would become separable. Hence, this quantum state is not entangled.  The proponents on the other hand said that what looks entangled is entangled. Stephen van Enk nicely summarized the discussion [31]. A major conceptual advance was made by Björk, Jonsson and Sanchez-Soto who proposed keeping the beams at the beam splitter output separate and interfering each of them with a local oscillator before detection [32, see also 33]. In the experiments, click detectors were used [34] but one could also use homodyne detection instead [35] or a combination of the two [36]. In view of the above, this measurement is exciting because it shows that the single photon splitting in the beam splitter leads to simultaneous signals in the two detectors – a correlation caused by the one single photon[2]. Furthermore, the transfer of single-photon entanglement in and out of an atomic system serving as a quantum memory has been demonstrated [37]. A Bell test was performed with this one particle state [38-39] and quantum teleportation was demonstrated [40-41]. It is interesting to note that the state $|\Psi\rangle = 2^{-1/2}(|01\rangle - |10\rangle)$ has a certain relation to the classical entanglement mentioned above, in that in both cases the relation between the mathematically non-separable mode function and the notion of quantum entanglement was discussed. It will be interesting to devise new experiments also in this context.

## Looking into complementarity using entanglement

Quantum optics allows for designing tests of complementarity [42]. In 2012 Menzel et al. published inspiring novel results [43] on the double slit experiment illuminated with one of the photons of an entangled pair. They studied experimentally and theoretically in how far the visibility of the interference fringes in the signal beam is affected by a which-slit measurement on the idler beam [43-44]. They emphasize that duality is well defined only for a single particle. Jakob and Bergou extended the concept of complementarity to bi-partite systems [45-46], which yet has to be applied to the two-photon double-slit experiment. Bolduc et al. pointed out that the auxiliary transverse variable provides an additional degree of freedom [47], which affects the interpretation of the experiment by Menzel et al. [43-44].

---

[2] Whatever this means in the quest for the physical reality of the quantum wave function. The least on can deduce is that the single photon is not local.

**The new experiments**
If new experiments can be done to further test a physical theory they should be done. The possibility to measure a single photon at several different locations opens a window for novel investigations The next steps exploring single photon entanglement seem obvious, with compelling goals:

(1) perform a delayed choice experiment with the $|\Psi\rangle = 2^{-1/2}(|01\rangle - |10\rangle)$ state using two simultaneous measurements.

(2) repeat the two-slit experiment with an entangled single photon.

(3) extend the studies to measurements at more than two points in space-time.

Instead of using the combination of a local oscillator and click detection, it may be advantageous to replace the click detector by a homodyne detector [35,36], or even better with an eight-port homodyne detector allowing for simultaneously measuring two conjugate field quadratures [48-50], keeping in mind that these different detectors correspond respectively to normal, symmetric and anti-normal ordering of the field operators and the associated quasi probability functions in phase space. For this task one can use integrated multiport splitters [51-53], which are interferometrically stable, making the proposed experiments viable. An interesting set-up uses a bi-prism [54], which may be combined with these different detector scenarios. Non-separable mode functions sometimes referred to as 'classical entanglement' [18-23] is closely related to single photon entanglement. The connection between these two areas and the potential synergy will be further explored.

**References**
[1]  N.D. Mermin, *arXiv*:quant-ph/**9609013** (1996)
[2]  N.D. Mermin, *Am. J. Phys.* **66**, 753 (1998)
[3]  P.C.W. Davies and J.R. Brown, 'The Ghost in the Atom – a dicussion oft he mysteries of quantum physics', Cambridge University Press (1986)
[4]  B.-G. Englert, *Eur. Phys. J. D* **67**, 238 (2013)
[5]  M.F. Pusey, J. Barrett and T. Rudolph, *Nature Phys.* **8**, 475 (2012); see also [54] p. 19
[6]  C.A. Fuchs, N.D. Mermin, R. Schack, *arXiv:***1311.5253** (2013); It is interesting to watch how a scientist having a strong opinion on which interpretation to prefer can convert to a different interpretation. The ambivalence in the scientific community may exist also in a single person. My experimentalist's soul is still waiting for things to check in the lab.
[7]  K.W. Murch, S.J. Weber, C. Macklin and I. Siddiqi, *Nature* **502**, 211 (2013)
[8]  J.A. Wheeler, *Annals New York Academy of Sciences* **655**, 349 (1992)
[9]  K.F. von Weizsäcker, *Zeitschrift für Physik* **118**, 489 (1941)
[10] J.S. Bell, *Physics (N.Y.)* **1**, 195 (1964)
[11] J.F. Clauser, M.A. Horne, M.A. Shimony and R.A. Holt, "Proposed Experiment to Test Local Hidden-Variable Theories", *Phys. Rev. Lett.* **23**, 880 (1969)
[12] A. Aspect, P. Grangier and G. Roger, *Phys. Rev. Lett.* **49**, 91 (1982)
[13] M. A. Rowe, D. Kielpinski, V. Meyer, C. A. Sackett, W. M. Itano, C. Monroe and D. J. Wineland, *Nature* **409**, 791 (2000)
[14] D.N. Matsukevich, P. Maunz, D.L. Moehring, S. Olmschenk and C.Monroe, *Phys. Rev. Lett.* **100**, 150404 (2008)
[15] M. Giustina, A. Mech, S. Ramelow, B. Wittmann, J. Kofler, J. Beyer, A. Lita, B. Calkins, T. Gerrits, S.W. Nam, R. Ursin and A. Zeilinger, *Nature* **497**, 227 (2013)
[16] B.G. Christensen, K.T. McCusker, J.B. Altepeter, B. Calkins, T. Gerrits, A.E. Lita, A. Miller, L.K. Shalm, Y. Zhang, S.W. Nam, N. Brunner, C.C.W. Lim, N. Gisin, P.G. Kwiat, *Phys. Rev. Lett.* **111**, 130406 (2013)

[17]  C. V. S. Borges, M. Hor-Meyll, J.A.O. Huguenin and A.Z. Khoury, *Phys. Rev. A* **82**, 033833 (2010)
[18]  R.J.C. Spreeuw, *Foundations of Physics* **28**, 361 (1998)
[19]  A. Holleczek, A. Aiello, Ch. Gabriel, Ch. Marquardt  and G. Leuchs, *arXiv:***1007.2528**  (2010)
[20]  X.-F. Qian and J.H. Eberly, *Opt. Lett.* **36**, 4110 (2011)
[21]  A. Holleczek, A. Aiello, Ch. Gabriel, Ch. Marquardt  and G. Leuchs, *Opt. Expr.* **19**, 9714 (2011)
[22]  F. Töppel, A. Aiello, Ch. Marquardt, E. Giacobino and G. Leuchs, *arXiv:***1401.1543** (2014)
[23]  L. Vaidman, Phys. Rev. A **87**, 052104 (2013)
[24]  J.A. Wheeler, in 'Quantum Theory and Measurement', edited by J.A. Wheeler and W.H. Zurek, pp.182-213, Princeton Univ. Press, Princeton, NJ (1984)
[25]  T. Hellmuth, H. Walther, A.G. Zajonc and W.P. Schleich, *Phys. Rev. A* **72**, 2533 (1987)
[26]  G. Weihs, Th. Jennewein, Ch. Simon, H. Weinfurter, and A. Zeilinger, *Phys. Rev. Lett.* **81**, 5039 (1998)
[27]  V. Jacques, E. Wu, F. Grosshans, F. Treussart, P. Grangier, A. Aspect and J.-F. Roch, *Science* **315**, 966 (2007)
[28]  A. Peruzzo, P. Shadbolt, N. Brunner, S. Popescu, and J.L. O'Brien, *Science* **338**, 634 (2012)
[29]  A. Danan, D. Farfurnik, S. Bar-Ad and L. Vaidman, *Phys. Rev. Lett.* **111**, 240402 (2013)
[30]  Z.-H. Li, M. Al-Amri and M.S. Zubairy, *Phys. Rev. A* **88**, 046102 (2013)
[31]  S.J. van Enk, *Phys. Rev. A* **72**, 064306 (2005)
[32]  G. Björk, P. Jonsson and L.L. Sanchez-Soto, *Phys. Rev. A* **64**, 042106 (2001)
[33]  S.M. Tan, D.F. Walls and M.J. Collett, *Phys. Rev. Lett.* **66**, 252 (1991)
[34]  B. Hessmo, P. Usachev, H. Heydari and G. Björk, *Phys. Rev. Lett.* **92**, 180401 (2004)
[35]  M. Fuwa, S. Takeda, M. Zwierz, H.M. Wiseman and A. Furusawa, arXiv:1412.7790 (2014
[36]  A.I. Lvovsky, H. Hansen, T. Aichele, O. Benson, J. Mlynek and S. Schiller, *Phys. Rev. Lett.* **87**, 050402 (2001)
[37]  K. S. Choi, H. Deng, J. Laurat and H. J. Kimble, *Nature* **452**, 67 (2008)
[38]  S.A. Babichev, J. Appel and A.I. Lvovsky, *Phys. Rev. Lett.* **92**, 193601 (2004)
[39]  M. D'Angelo, A. Zavatta, V. Parigi and M. Bellini, *Phys. Rev. A* **74**, 052114 (2006)
[40]  S. Giacomini, F. Sciarrino, E. Lombardi and F. De Martini, *Phys. Rev. A* **66**, 030302 (2002)
[41]  H. de Riedmatten, I. Marcikic, W. Tittel, H. Zbinden, D. Collins and N. Gisin, *Phys. Rev. Lett.* **92**, 047904 (2004)
[42]  M.O. Scully, B.-G. Englert and H. Walther, *Nature* **351**, 111 (1991)
[43]  R. Menzel, A. Heuer, D. Puhlmann and W.P. Schleich, *Proc. Natl. Acad. Sci. (U.S.A.)* **109**, 9314 (2012)
[44]  R. Menzel, A. Heuer, D. Puhlmann, K. Dechoum, M. Hillery, M.J.A. Spähn and W.P. Schleich, *J. Mod. Opt.* **60**, 86 (2013)
[45]  M. Jakob and J.A. Bergou, *Phys. Rev. A* **76**, 052107 (2007)
[46]  M. Jakob and J.A. Bergou, *Opt. Commun.* **283**, 827 (2010)
[47]  E. Bolduc, J. Leach, F.M. Miatto, G. Leuchs and R.W. Boyd, *arXiv:***1402.6487** (2012)
[48]  M.J. Collett, R. Loudon and C.W. Gardiner, *J. Mod. Opt.* **34**, 881, (1987)
[49]  M. Freyberger, K. Vogel and W.P. Schleich,  *Phys. Lett. A* **176**, 41 (1993); see also W.P. Schleich, "Quantum optics in phase space", Wiley-VCH (2001)
[50]  W. Vogel and D.-G. Welsch, "Quantum optics", John Wiley & Sons (2006)
[51]  R. Ulrich and G. Ankele, *Appl. Phys. Lett.* **27**, 337 (1975)
[52]  P. Roth and G. Voirin, *Proc. SPIE* **1014**, *Micro-Optics* 35 (1989); doi:10.1117/12.949392
[53]  A. Peruzzo, A. Laing, A. Politi, T. Rudolph and J.L. O'Brien,  *arXiv:* **1205.4926** (2012)
[54]  V. Jacques, E. Wu, T. Toury, F. Treussart, A. Aspect, P. Grangier, and J.-F. Roch, *Eur. Phys. J. D* **35** , 561 (2005)
[55]  P. Wallden, V. Dunjko and E. Andersson, *J. Phys. A: Math. Theor.* **47**, 125303 (2014)